\documentclass{PoS}
\usepackage{amsmath,amssymb}
\usepackage{bm}
\usepackage{graphicx}

\title{More about vacuum structure of Linear Sigma Model}

\ShortTitle{More about vacuum structure of Linear Sigma Mode}

\author{\speaker{Tomomi Sato}\\
        High Energy Accelerator Research Organization (KEK), %
        Tsukuba 305-0801, Japan\\
        Graduate University for Advanced Studies (SOKENDAI), %
        Tsukuba 305-0801, Japan\\
        E-mail: \email{tomomis@post.kek.jp}}

\author{Norikazu Yamada\\
        High Energy Accelerator Research Organization (KEK), %
        Tsukuba 305-0801, Japan\\
        Graduate University for Advanced Studies (SOKENDAI), %
        Tsukuba 305-0801, Japan\\
        E-mail: \email{norikazu.yamada@kek.jp}}

\abstract{
In the study of critical phenomena of QCD, a linear sigma model
(LSM) is often analyzed as it shares many properties with QCD.
Motivated by recent arguments on effective restoration of the $U_A(1)$
symmetry around the critical temperature, the renormalization group flow
of $U(2)\otimes U(2)$ LSM with a small violation of the $U_A(1)$
symmetry is examined in the traditional $epsilon$ expansion in three
dimensions.
With a mass-dependent renormalization scheme, we investigate the
attractive basin flowing into the $O(4)$ LSM in the parameter space and
its dependence on the size of the $U_A(1)$ breaking.
Special emphasis is put on how the decoupling of the heavier degrees of
freedom occur as approaching the $O(4)$ LSM.
}

\FullConference{31st International Symposium on Lattice Field Theory - LATTICE 2013\\
		July 29 - August 3, 2013\\
		Mainz, Germany}

\begin{document}

\section{Introduction}
\label{sec:introduction}

One of the most important features of QCD is the chiral phase
transition at a finite temperature.
Thirty years ago, Pisarski and Wilczek revisited the known results of
the $\epsilon$ expansion in a linear sigma model and pointed out that
the chiral phase transition of massless $N_f$-flavor QCD is first order for
$N_f\ge 3$ and, importantly, that the effective restoration of $U_A(1)$
symmetry is directly relevant in the determination of the nature of the
chiral phase transition in two-flavor QCD~\cite{Pisarski:1983ms}.
$U_A(1)$ symmetry is broken by quantum anomaly at the quark-gluon (or
Lagrangian) level, independently of occurrence of spontaneous chiral
symmetry breaking (S$\chi$SB).
It is, then, naively expected that the breaking effect should be taken
over by dynamics at the hadron level.
Indeed, a relatively large mass difference observed between $\eta$ and
$\eta'$ mesons at zero temperature provides an experimental support.
However, analytical and numerical studies about the $U_A(1)$ symmetry
breaking suggest that the breaking effects are extremely suppressed in
physical observables around the critical temperature ($T_c$) for
S$\chi$SB~\cite{Bazavov:2012qja,Aoki:2012yj,Cossu:2013uua,Buchoff:2013nra}.
Thus, $U_A(1)$ symmetry appears to restore, at least, approximately
around $T_c$.

In this study, we focus on the two-flavor QCD.
In order to gain qualitative information about the chiral transition, a
linear sigma model (LSM) has been studied as it is considered to be an
effective theory describing the critical phenomena of QCD.
When the effect of $U_A(1)$ breaking on the system is sufficiently
large, $U(2)\otimes U(2)$ symmetry is explicitly broken down to $O(4)$.
The renormalization group (RG) flow of $O(4)$ LSM is known to possess a
stable infrared fixed point (IRFP), which indicates that the chiral
phase transition in two-flavor QCD is second order governed by the
critical exponents in the $O(4)$ [or $O(4)/O(3)$] universality class.

On the other hand, when $U_A(1)$ symmetry is effectively restored around
$T_c$, the transition should be analyzed in $U(2)\otimes U(2)$ LSM,
which has two independent couplings.
Recently, Pelissetto and Vicari confirmed the existence of the IRFP in
$U(2)\otimes U(2)$ LSM at the five- or six-loop
order~\cite{Pelissetto:2013hqa}, which suggests that the chiral
transition is second but following the $U(2)\otimes U(2)/U(2)$ scaling.
Therefore, the size of $U_A(1)$ breaking at $T_c$ is the key to
discriminate the nature of the transition in two-flavor QCD.
We explore how the transition goes through when the $U_A(1)$ breaking is
small but finite.
Especially, with a mass dependent renormalization scheme, we see how the
decoupling of massive degrees of freedom occurs along the flow toward
the infrared limit.
Similar analysis with $c_A=0$ is performed in Ref.~\cite{Aoki:2012LSM}.

\section{Pisarski and Wilczek's argument}
\label{sigma}

In this section, we review the Pisarski and Wilczek's argument about
chiral phase transition\cite{Pisarski:1983ms}.
We introduce an $N_f \times N_f$ complex matrix field $\Phi$ as
\begin{align}
  \Phi
 =\sqrt{2}(\sigma+i\eta)T^0+\sqrt{2}(\delta_i+i\pi_i)T^i
 =\sqrt{2}(\phi_0-i\chi_0)T^0+\sqrt{2}(\chi_i+i\phi_i)T^i,
 \label{eq:Phi}
\end{align}
where $\phi_a=\{\sigma,\pi_i\}$, $\chi_a=\{-\eta,\delta_i\}$
$(a=0,1,\cdots,N_f)$, $T^0=\bm{1}/2$, and $T^i$ is the generator of
$SU(N_f)$ group.
Then, let $\Phi$ transform as
\begin{align*}
 \Phi \to e^{2i\theta_A}L^{\dagger}\Phi R\ \
 (L\in SU_L(N_f),\ \ R\in SU_R(N_f)),
\end{align*}
under chiral and $U_A(1)$ transformations.\footnote{Appearance of phase
$\theta_A$ comes from the $U_A(1)$ transformation.
$U_V(1)$ symmetry is irrelevant in the following discussion and hence is
omitted.}
Most general renormalizable Lagrangian conserving chiral and $U_A(1)$
rotations is
\begin{align}
\label{U(N)}
   \mathcal{L}_0
 = \frac{1}{2}\mathrm{tr}\left[\partial_{\mu}\Phi^{\dagger}
   \partial^{\mu}\Phi \right]
  +\frac{1}{2}m^2\mathrm{tr}\left[\Phi^{\dagger}\Phi \right]
  +\frac{\pi^2}{3}g_1\left(\mathrm{tr}[\Phi^{\dagger}\Phi]\right)^2
  +\frac{\pi^2}{3}g_2\mathrm{tr}\left[(\Phi^{\dagger}\Phi)^2\right].
\end{align}
Because $\Phi$ corresponds to the order parameter of chiral symmetry,
nonzero vev of $\Phi$ indicates S$\chi$SB.
By assuming that this system ends up with the second order phase
transition, the system near $T_c$ should be well described by a three
dimensional field theory  because of the divergent correlation length.
Then, the temperature dependent mass $m(T)$ vanishes at $T_c$.
Following the traditional $\epsilon$ expansion approach, the $\beta$
functions are calculated.
If the assumption of the second order is right, the system will show an
stable infrared fixed point (IRFP).

At the one-loop level, it turns out that for $N_f>\sqrt{3}$ there is no
IRFP in this system and all couplings flow into the region
\begin{align}
 N_fg_1+g_2\geqq 0,\ \ g_1+g_2\geqq 0,
\end{align}
where the effective potential is not bounded below at the tree level.
With this result, it is expected that the chiral transition of
$N_f$-flavor QCD with $N_f\geqq 2$ is fluctuation induced first
order.\footnote{
Pelissetto and Vicari found an IRFP after the higher orde calculations
and the resummation~\cite{Pelissetto:2013hqa}.
}

Next, the case with the $U_A(1)$ breaking is examined.
More precisely, the following $U_A(1)$ breaking terms are added to
Lagrangian, eq.~(\ref{U(N)}),
\begin{align*}
   \mathcal{L}_A
 \equiv
   -\frac{c_A}{4}(\mathrm{det}\,\Phi+\mathrm{det}\,\Phi^{\dagger})
\end{align*}
for $N_f\geqq 3$, and
\begin{align}
\label{anom}
    \mathcal{L}_A
 \equiv
  - \frac{c_A}{4}(\mathrm{det}\,\Phi+\mathrm{det}\,\Phi^{\dagger})
  + \frac{\pi^2}{3}x\, \mathrm{Tr}[\Phi\Phi^{\dagger}]
    (\mathrm{det}\,\Phi+\mathrm{det}\,\Phi^{\dagger})
  + \frac{\pi^2}{3}y\,
    (\mathrm{det}\,\Phi+\mathrm{det}\,\Phi^{\dagger})^2
\end{align}
for $N_f=2$.\footnote{The term
$\mathrm{det}\Phi\,\mathrm{det}\Phi^\dagger$ is not independent of the
other terms.}

Hereafter, we focus on the two-flavor QCD.
Rewriting the total Lagrangian, $\mathcal{L}_0+\mathcal{L}_A$, in terms
of the component fields, we obtain
\begin{align}
\label{Lfull}
 \mathcal{L}=&\frac{1}{2}(\partial_{\mu}\phi_a)^2+\frac{1}{2}(\partial_{\mu}\chi_a)^2
 +\frac{m_{\phi}(T)^2}{2}{\phi_a}^2+\frac{m_{\chi}(T)^2}{2}{\chi_a}^2
\notag \\
 &+\frac{\pi^2}{3}\left[
 \lambda({\phi_a}^2)^2+(\lambda-2x)({\chi_a}^2)^2
 +2(\lambda+g_2-z){\phi_a}^2{\chi_b}^2
 -2g_2(\phi_a\chi_a)^2\right],
\end{align}
where $\lambda=g_1+g_2/2+x+y$ and $z=x+2y$.
It is important to notice that the $c_A$ term in eq.~(\ref{anom}) makes
the degeneracy between $\phi_a$ and $\chi_a$ separate off as
\begin{align}
 \label{eq:mass2}
 m_{\phi}^2=m(T)^2-\frac{c_A}{2},\ \ \
 m_{\chi}^2=m(T)^2+\frac{c_A}{2}.
\end{align}
Naively, one expects that $\chi$'s decouple from the system and hence
Lagrangian is reduced to $O(4)$ LSM
\begin{align}
\label{O(4)}
  \mathcal{L}
=&\frac{1}{2}(\partial_{\mu}\phi_a)^2
 +\frac{1}{2} \left(m(T)^2-\frac{c_A}{2}\right)\phi_a^2
 +\frac{\pi^2}{3}\lambda (\phi_a^2)^2.
\end{align}
$O(4)$ LSM is known to have a stable IRFP and hence the phase transition
is expected to be second order.
Thus, in the two-flavor QCD with sufficiently large $c_A$, the chiral
phase transition is expected to be second order in the $O(4)$
universality class.

\section{RG flow and Decoupling}
\label{main}

We revisit the latter case in sec.~\ref{sigma}, the case with a finite
$U_A(1)$ breaking, with special emphasis on how the effects of the
massive degrees of freedom ($\chi$'s) disappear.
In this case, the critical temperature $T_c$ is defined so that the mass
of the lightest degrees of freedom vanish as follows,
\begin{align*}
 m^2_{\phi}(T_c)=m^2(T_c)-\frac{c_A}{2}=0,
\ \
 m^2_{\chi}(T_c)=m^2(T_c)+\frac{c_A}{2}=c_A.
\end{align*}
Thus now $\chi$ has a finite mass.
In the calculation of the $\beta$ function, we simply follow the
traditional $\epsilon$ expansion method, where after the dimensional
regularization with $D=4-\epsilon$ $\epsilon$ is set to unity.
In order to trace the effects of the massive fields, we take a mass
dependent renormalization scheme by imposing the following
renormalization conditions to four-point functions,
\begin{align}
  G_4(\phi_1(p_1),\phi_1(p_2);\phi_2(p_3)\phi_2(p_4))|_{s=t=u=-\mu^2}
 &=-i\frac{8}{3}\pi^2\lambda\prod_i \frac{i}{p_i^2-m_i^2},
\\
 G_4(\eta_1(p_1),\eta_1(p_2);\eta_2(p_3)\eta_2(p_4))|_{s=t=u=-\mu^2}
 &=-i\frac{8}{3}\pi^2(\lambda-2x)\prod_i \frac{i}{p_i^2-m_i^2},
\\
 G_4(\phi_1(p_1),\eta_2(p_2);\phi_1(p_3)\eta_2(p_4))|_{s=t=u=-\mu^2}
 &=-i\frac{4}{3}\pi^2(\lambda+g_2-z)\prod_i \frac{i}{p_i^2-m_i^2},
\\
 G_4(\phi_1(p_1),\eta_2(p_2);\phi_2(p_3)\eta_1(p_4))|_{s=t=u=-\mu^2}
 &=i\frac{4}{3}\pi^2g_2\prod_i \frac{i}{p_i^2-m_i^2}.
\end{align}
Then, we obtain the following $\beta$ functions to the one loop and the
first order in $\epsilon$,
\begin{align}
\label{bl}
   \beta_{\lambda}
 \equiv
   \mu\frac{d\lambda}{d\mu}
 = &-\epsilon\lambda+2\lambda^2
    +\frac{1}{6}f(\hat\mu)
     (4\lambda^2+6\lambda g_2+3g_2^2-8\lambda z-6g_2z+4z^2),
\\
\label{bg2}
   \beta_{g_2}
 \equiv
   \mu\frac{dg_2}{d\mu}
 =&-\epsilon g_2+\frac{1}{3}\lambda g_2
   +\frac{1}{3}f(\hat\mu)g_2
    (\lambda-2x)+\frac{1}{3}h(\hat\mu)g_2(4\lambda+g_2-4z),
\\
\label{bx}
   \beta_x
 \equiv
   \mu\frac{dx}{d\mu}
  =&-\epsilon x+4f(\hat\mu)(\lambda x-x^2)
\notag \\
   &+\frac{1}{12}(1-f(\hat\mu))
    (8\lambda^2-6\lambda g_2-3g_2^2+8\lambda z+6g_2z-4z^2-12xz),
\\
\label{bz}
   \beta_z
 \equiv
   \mu\frac{dz}{d\mu}
  =&-\epsilon z+\frac{1}{2}(2\lambda^2-\lambda g_2+2\lambda z)
    -\frac{1}{6}h(\hat\mu)(4\lambda^2+3g_2^2-8\lambda z+4z^2)
\notag \\
   &+\frac{1}{6}f(\hat\mu)(-2\lambda^2+3\lambda g_2+3g_2^2
    -2\lambda z-6g_2 z+12\lambda x+6g_2x+4z^2),
\end{align}
where $\hat\mu\equiv\mu/\sqrt{c_A}$
and
\begin{align}
   f(x)
 =1-\frac{4}{x\sqrt{4+x^2}}\arctan \sqrt{\frac{x^2}{4+x^2}},
\ \ \
   h(x)
 =1-\frac{1}{x^2}\log [1+x^2]
\end{align}
with $f(0)=g(0)=0$ and
$\lim_{x\rightarrow \infty}f(x)=\lim_{x\rightarrow \infty}h(x)=1$.

When $c_A$ is zero, these $\beta$ functions are consistent with
what derived by Aoki, Fukaya and Taniguchi~\cite{Aoki:2012LSM}, 
and the one-loop analysis does not lead to the existence
of the stable IRFP.
In the following, we only consider the case with a finite $c_A$.

\begin{figure}[tbp]
 \begin{center}
  \includegraphics[width=0.6 \textwidth]{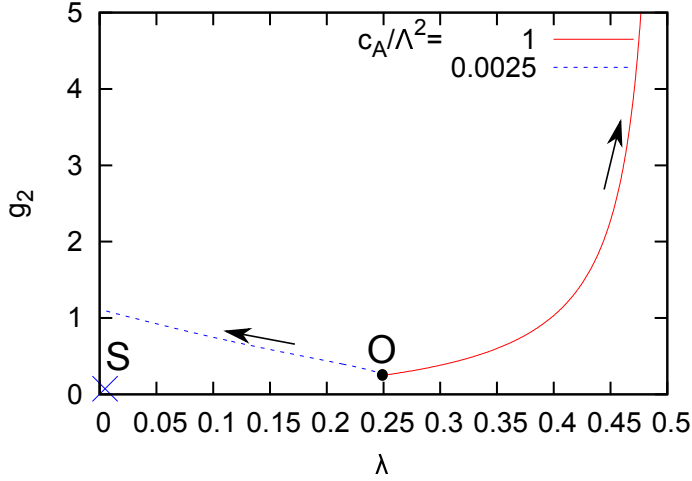}
 \vspace{-2ex}
 \end{center}
 \caption{
 RG flows starting with the initial location {\bf O} corresponding to
 $(\lambda(\Lambda),\ g_2(\Lambda),\ x(\Lambda),\ z(\Lambda))=
  (0.25,\ 0.25,\ 0,\ 0)$ and with two different values of $c_A$.
 $\epsilon$ is set to 1.
 The IRFP of $U(2)\otimes U(2)$ LSM without $U_A(1)$ breakings
 employing a different scheme with higher order corrections is denoted
 by {\bf S}, $(\lambda,\ g_2)\sim$ (0.0048,0.073) as a
 reference~\cite{Pelissetto:2013hqa}.
 }
 \label{fig:flow}
 \vspace{-2ex}
\end{figure}

\begin{figure}[tbp]
 \begin{center}
  \includegraphics[width=0.9 \textwidth]{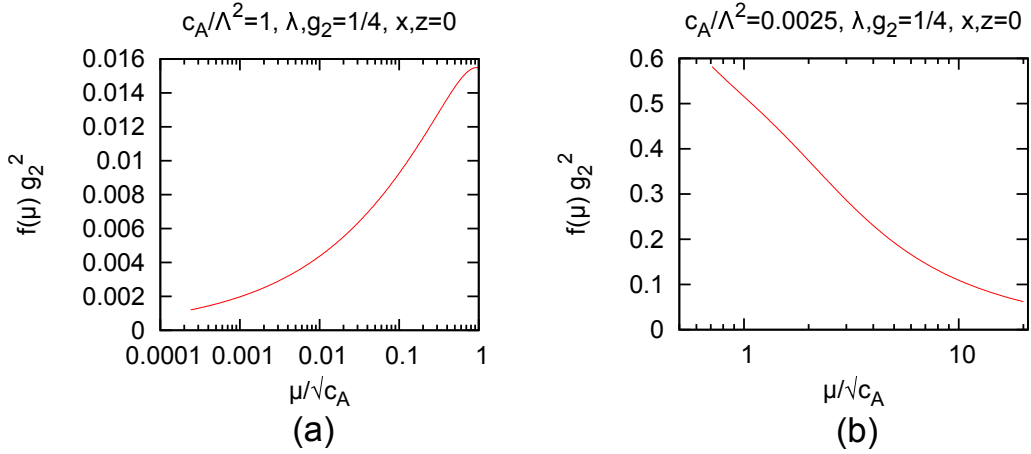}
  \vspace{-5ex}
 \end{center}
 \caption{Contribution of the $g_2^2$ term to $\beta_{\lambda}$ with (a)
 $c_A/\Lambda^2=1$ and (b) $c_A/\Lambda^2=0.0025$.}
 \label{fig:cont}
 \vspace{-2ex}
\end{figure}

Let $\Lambda$ be the initial energy scale, at which the theory is defined.
As the initial condition, we take
$(\lambda(\Lambda),\ g_2(\Lambda),\ x(\Lambda),\ z(\Lambda))
=(0.25,\ 0.25,\ 0,\ 0)$ as an example.
By numerical calculations, we obtain, at least, two types of RG flow as
shown in Fig.~\ref{fig:flow}.
To see how the flow depends on the size of the $U_A(1)$ breaking, two
different values of $c_A$ are tested, $c_A/\Lambda^2=1$ and 0.01.
On the flow with the lager $c_A$, $c_A/\Lambda^2=1$ (red solid curve),
$\lambda$ converges to $\epsilon/2$, which is the IRFP of $O(4)$ LSM,
eq.~(\ref{O(4)}).
It is interesting to note that, as approaching the $O(4)$ fixed point,
both $g_2$ and $z$ diverge.
These are the couplings of the ${\phi_a}^2{\chi_b}^2$ and
$(\phi_a\chi_a)^2$ terms, so the decoupling of $\chi$
appears to be non-trivial.
In order to see what happens, we look at the IR limit of the flow in the
vicinity of $\lambda=\epsilon/2$.
In this limit, the $\beta$ functions, eqs.~(\ref{bg2}) and (\ref{bz}),
can be approximated to be
\begin{align*}
 \beta_{g_2}(\mu\to 0)&\sim -\frac{5}{6}\epsilon g_2
 +O\left(\frac{\mu}{\sqrt{c_A}}\right),
\\
 \beta_z(\mu\to 0)&\sim-\frac{1}{2}\epsilon z-\frac{1}{4}\epsilon g_2+\frac{\epsilon^2}{4}
 +O\left(\frac{\mu}{\sqrt{c_A}}\right).
\end{align*}
By solving these, we obtain the divergent behaviors,
\begin{align*}
      g_2(\mu)
 \sim g_2(\Lambda)\left(\frac{\mu}{\Lambda}\right)^{-\frac{5}{6}\epsilon},
\ \ \
      z(\mu)
 \sim g_2(\mu).
\end{align*}
But the contributions of these terms to $\beta_{\lambda}$ vanish in the
IR limit as shown in Fig.~\ref{fig:cont}(a) because
\begin{align*}
 f(\hat\mu)\, g_2^2
 \sim
 \mu^{2-\frac{5}{3}\epsilon}.
\end{align*}
Eventually, $\beta_{\lambda}$ reduces to the $\beta$ function of the $O(4)$ LSM,
\begin{align}
\label{bO(4)}
 \beta_{O(4)}=-\epsilon\lambda+2\lambda^2.
\end{align}
Thus, on the flow approaching the $O(4)$ LSM, the massive fields
$\chi_a$ indeed decouples from the system as expected in the
general argument on the decoupling~\cite{Appelquist:1974tg}.

The other case in Fig.~\ref{fig:flow} (blue dashed curve) describes the
RG flow with the smaller $c_A$ ($c_A/\Lambda^2=0.0025$).
In this case, $\lambda$ flows into the negative region and eventually
diverges (within the one-loop analysis).
Furthermore, the contribution of the $g^2_2$ term to $\beta_{\lambda}$
also diverges towards the IR limit as shown in Fig.~\ref{fig:cont}(b).
Thus, the one-loop analysis tells that, even if $\mu$ is much smaller
than the mass of $\chi_a$'s, the presence of $\chi_a$ fields gives
significant effects to the running of the couplings.
There are two possibilities.
The first one is that (after including higher order corrections) the
flow eventually goes into another stable IRFP and then the terms in the
$\beta$ functions stop growing.
The second is that the assumption of the second order transition is
wrong.

\begin{figure}[tb]
 \begin{center}
  \includegraphics[width=0.9 \textwidth]{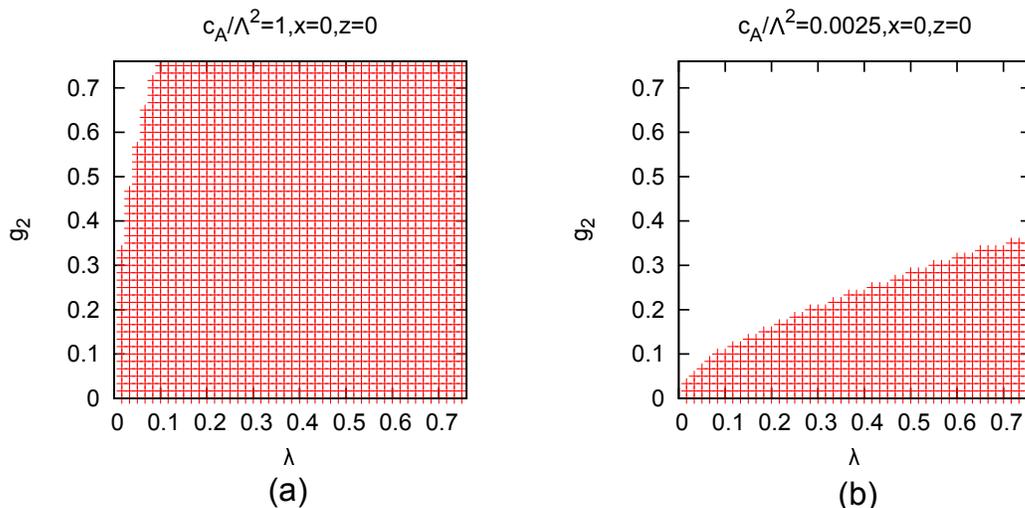}
  \vspace{-5ex}
 \end{center}
 \caption{The attractive basin in two dimensional space spanned by the
 initial couplings, $\lambda(\Lambda)$ and $g_2(\Lambda)$ (red region)
 is shown, where $x(\Lambda)=z(\Lambda)=0$ and $\epsilon=1$.
 Two different cases are plotted; (a) $c_A/\Lambda^2=1$ and (b)
 $c_A/\Lambda^2=0.0025$.}
 \label{fig:reg1}
 \vspace{-2ex}
\end{figure}

We determine the attractive basin in the initial coupling space, with
which the system flows into $O(4)$ LSM.
The result is shown by the red region of Fig.~\ref{fig:reg1}.
It is seen that the attractive basin expands with $c_A$ as naively
expected.

\section{Summary}

With the leading order calculation of the $\epsilon$ expansion and a
mass dependent renormalization scheme, we examined the RG flow of
$U(2)\otimes U(2)$ LSM with the $U_A(1)$ breaking.
It is found that the flow depends on the size of the $U_A(1)$ breaking
(or $c_A$) and the attractive basin flowing into $O(4)$ LSM expands with
$c_A$ as naively expected.
Starting from the outside of the basin, the decoupling of the massive
fields does not occur and the transition is likely to be first order.
Of course, the flow is subject to the higher order corrections in the
expansion, and such a calculation is clearly necessary.

In this analysis, we only study the flow starting from a fixed initial
condition
$(\lambda(\Lambda),\ g_2(\Lambda),$ $x(\Lambda),\ z(\Lambda))=
 (0.25,\ 0.25,\ 0,\ 0)$.
An extension to the case with other initial conditions is
straightforward, and such a study is on going to see the general feature
of this system.

\section*{Acknowledgements}
This work is supported in part by JSPS KAKENHI 22740183.

\end{document}